\documentclass[,
nofootinbib,
twocolumn,
superscriptaddress,
amsmath,amssymb,
aps,
prl,
prstper,
floatfix,
]{revtex4-1}
\usepackage{mathrsfs}
\usepackage[pdftex]{graphicx}% Include figure files
\usepackage{dcolumn}% Align table columns on decimal point
\usepackage{bm}% bold math
\usepackage{longtable}
\usepackage{color,hyperref}
\usepackage{tikz}
\usepackage{longtable}
\hypersetup{colorlinks,breaklinks,linkcolor=blue,urlcolor=blue,anchorcolor=blue,citecolor=blue}
\usepackage{comment}
\usepackage{multirow}            
            
\newcolumntype{d}[1]{D{.}{.}{#1}}

\newcommand{\ra}[1]{\renewcommand{\arraystretch}{#1}}
	\newcommand{\ket}[1]{\left| #1 \right\rangle}
	\newcommand{\bra}[1]{\left\langle #1 \right|}

\newcommand{\reduceE}[2]{\langle #1||M(E2)||#2\rangle}
\newcommand{\element}[2]{$^{#1}$\textrm{#2}}

\begin{document}
\title{Shape coexistence and the role of axial asymmetry in $^{72}$Ge}

\author{A. D. Ayangeakaa}
 \email{ayangeakaa@anl.gov}
 \affiliation{Physics Division, Argonne National Laboratory, Argonne, Illinois 60439, USA}
 \author{R. V. F. Janssens}
\affiliation{Physics Division, Argonne National Laboratory, Argonne, Illinois 60439, USA}
\author{C. Y. Wu} 
\affiliation{Lawrence Livermore National Laboratory, Livermore, California 94550, USA}
\author{J. M. Allmond}
 \affiliation{Physics Division, Oak Ridge National Laboratory, Oak Ridge, Tennessee 37831, USA}
 \author{J. L. Wood}
\affiliation{School of Physics, Georgia Institute of Technology, Atlanta, Georgia 30332, USA}
\author{S. Zhu}
\affiliation{Physics Division, Argonne National Laboratory, Argonne, Illinois 60439, USA}
  \author{M. Albers}
\altaffiliation[Present Address: ]{Ernst \& Young GmbH, Mergenthalerallee 3-5, D-65760 Eschborn, Germany.}
\affiliation{Physics Division, Argonne National Laboratory, Argonne, Illinois 60439, USA}

\author{S. Almaraz-Calderon}
\altaffiliation[Present Address: ]{Department of Physics, Florida State University, Tallahassee, Florida 32306, USA.}
\affiliation{Physics Division, Argonne National Laboratory, Argonne, Illinois 60439, USA}
\author{B. Bucher} 
\affiliation{Lawrence Livermore National Laboratory, Livermore, California 94550, USA}
\author{M. P. Carpenter}
\affiliation{Physics Division, Argonne National Laboratory, Argonne, Illinois 60439, USA}
\author{C. J. Chiara}
\altaffiliation[Present Address: ]{U.S. Army Research Laboratory, Adelphi, Maryland 20783, USA.}
\affiliation{Physics Division, Argonne National Laboratory, Argonne, Illinois 60439, USA}
\affiliation{Department of Chemistry and Biochemistry, University of Maryland, College Park, Maryland 20742, USA}

\author{D. Cline}
\affiliation{Department of Physics and Astronomy, University of Rochester, Rochester, New York 14627, USA}

\author{H. L. Crawford}
\altaffiliation[Present Address: ]{Nuclear Science Division, Lawrence Berkeley National Laboratory, Berkeley, California 94720, USA.}
\affiliation{Institute of Nuclear and Particle Physics, Department of Physics and Astronomy, Ohio University, Athens, Ohio 45701, USA}
\author{H. M. David}
\altaffiliation[Present Address: ]{GSI, Helmholtzzentrum f\"ur Schwerionenforschung GmbH, 64291 Darmstadt, Germany.}
\affiliation{Physics Division, Argonne National Laboratory, Argonne, Illinois 60439, USA}

\author{J. Harker}
 \affiliation{Physics Division, Argonne National Laboratory, Argonne, Illinois 60439, USA}
\affiliation{Department of Chemistry and Biochemistry, University of Maryland, College Park, Maryland 20742, USA}

\author{A. B. Hayes}
\affiliation{Department of Physics and Astronomy, University of Rochester, Rochester, New York 14627, USA}

\author{C. R. Hoffman}
\affiliation{Physics Division, Argonne National Laboratory, Argonne, Illinois 60439, USA}

\author{B. P. Kay}
\affiliation{Physics Division, Argonne National Laboratory, Argonne, Illinois 60439, USA}

\author{K. Kolos}
\affiliation{Department of Physics and Astronomy, University of Tennessee, Knoxville, Tennessee 37996, USA}

\author{A. Korichi}
\affiliation{Centre de Spectrom{\'e}trie Nucl{\'e}aire et de Spectrom{\'e}trie de Masse CSNSM, CNRS/IN2P3 and  
Universit{\'e} Paris-Sud, F-91405 Orsay Campus, France}

\author{T. Lauritsen}
\affiliation{Physics Division, Argonne National Laboratory, Argonne, Illinois 60439, USA}

\author{A. O. Macchiavelli}
\affiliation{Nuclear Science Division, Lawrence Berkeley National Laboratory, Berkeley, California 94720, USA}

\author{A. Richard}
\affiliation{Institute of Nuclear and Particle Physics, Department of Physics and Astronomy, Ohio University, Athens, Ohio 45701, USA}

\author{D. Seweryniak}
\affiliation{Physics Division, Argonne National Laboratory, Argonne, Illinois 60439, USA}

\author{A. Wiens}
\affiliation{Nuclear Science Division, Lawrence Berkeley National Laboratory, Berkeley, California 94720, USA}

\date{\today}

\begin{abstract} 
The quadrupole collectivity of low-lying states and the anomalous behavior of the $0^+_2$ and $2^+_3$ levels in $^{72}$Ge are investigated via projectile multi-step Coulomb excitation with GRETINA and CHICO-2. A total of forty six  $E2$ and $M1$ matrix elements connecting  fourteen low-lying levels were determined using the least-squares search code, \textsc{gosia}. Evidence for triaxiality and shape coexistence, based on the model-independent shape invariants deduced from the Kumar-Cline sum rule, is presented. These are interpreted using a simple two-state mixing model as well as multi-state mixing calculations carried out within the framework of the triaxial rotor model. The results represent a significant milestone towards the understanding of the unusual structure of this nucleus. 
\end{abstract}
\pacs{}
\maketitle

The structure of low-lying states in even-even Ge isotopes has been the subject of intense scrutiny for many years due to the inherent challenge of interpreting their systematics as a function of mass $A$. These nuclei possess at least one excited $0^+$ state in their low-energy spectrum that differs from the ground state in its properties. Systematically, the energy of the $0^+_2$ level varies parabolically with $A$ and reaches a minimum in $^{72}$Ge, where it becomes the first excited state. The existence of even-mass nuclei with a $J^\pi= 0^+$ first excited state is an uncommon phenomenon which, to date, has been observed in only a few nuclei located near or at closed shells: $^{16}$O~\cite{PhysRev.101.254}, $^{40}$Ca~\cite{BROWN196687,Brown1966401}, $^{68}$Ni~\cite{Bernas1982279,PhysRevC.89.021301}, $^{90}$Zr~\cite{PhysRevC.4.196}, $^{180,182}$Hg~\cite{PhysRevC.84.034307Elseviers,PhysRevC.84.034308,PhysRevC.50.2768}, $^{184,186,188,190,192,194}$Pb~\cite{refId0,Andreyev2000,Bijnens1996,VANDUPPEN1985354,PhysRevC.35.1861,PhysRevLett.52.1974}. There are also examples of such nuclei where a subshell appears to play a role similar to a closed shell such as $^{96,98}$Zr~\cite{Lahanas1986399,Fogelberg1971372}. These cases have all been explained as resulting from shape coexistence due to the presence of intruder configurations; {\it i.e.}, configurations involving the excitation of at least one pair of nucleons across a shell or subshell energy gap~\cite{RevModPhys.83.1467}. 

The structure of $^{72}$Ge is highly unusual in that this nucleus is far from closed shells and, yet, possesses a $0^+$ first excited state. It shares this distinction with only two other known nuclei: $^{72}$Kr~\cite{Varley1987463} and $^{98}$Mo~\cite{Zielin"ska20023}. It should be noted that while $^{68}$Ni might be doubly magic~\cite{Bernas1982279,PhysRevC.88.041302} with a presumed subshell closure at $N=40$~\cite{PhysRevLett.74.868}, there is no evidence to date of spherical ground-state configurations in any of the other $N=40$ isotones~\cite{PhysRevLett.83.3613}. There are, however, strong experimental indications of enhanced collective behavior in or near the ground states of the $N=40$ neutron-rich Fe and Cr isotones~\cite{PhysRevC.81.051304}. In the Ge isotopes, the absence of a subshell closure at $N=40$ can be inferred from the fact that the $2^+_1$ level with the highest energy appears in $^{70}$Ge rather than $^{72}$Ge. Thus, understanding the nature and origin of this anomalous $0_2^+$ state in $^{72}$Ge has been a major challenge for collective model descriptions.

The theory of collectivity in nuclei is predominantly focused on models with quadrupole degrees of freedom. The simplest of these models, based on the quantization of a liquid drop~\cite{bohr1998nuclear,rowe2010fundamentals}, describes quadrupole collectivity as either due to harmonic quadrupole vibrations of a spherical shape or to rotations and vibrations of a deformed quadrupole shape: a prolate or oblate spheroid or an axially asymmetric ellipsoid. None of these variants of nuclear collectivity possesses $0^+$ first excited states. More sophisticated models abound, for example models based on boson degrees of freedom~\cite{klein-RevModPhys.63.375}, and most notably the interacting boson model~\cite{iachello1987interacting}, but to produce a first excited $0^+$ state usually requires extreme parameter choices fitted to the nucleus under investigation. Other models with quadrupole collectivity introduce ``pair-excitations" in an \emph{ad hoc} way. These approaches are motivated by the shape coexistence phenomena observed at and near closed shells. These models contain parameters that are fitted to the properties of low-lying excited $0^+$ states, most notably to a large pair excitation energy parameter which is overcome by enhanced quadrupole correlations. The latter lower the energy of the pair configuration, even to the extent that the intruder configuration becomes the ground state~\cite{Iwasaki01101976,Duval1981223,Duval1983297,prochniack2005}. It should be noted that microscopic collective models are now beginning to provide first insights into such structures~\cite{PhysRevC.80.014305}.

In this letter, a detailed study of quadrupole collectivity in $^{72}$Ge by multi-step Coulomb excitation is reported. This nucleus has been studied by Coulomb excitation in the earlier work of Kotli{\'n}ski {\it et al.}~\cite{Kot1990646}. The extended set of $E2$ matrix elements obtained in the present study now permits a model-independent view of the shapes exhibited by $^{72}$Ge up to moderate spin (8$\hbar$). This, to the best of our knowledge, is the most extensive set of shape invariants achieved for any nucleus with a first-excited $0^+$ state. The deformation and asymmetry of the ground-state band are found to be remarkably constant. The shape invariants from the yrast and non-yrast states support an interpretation based on shape coexistence, with the ground-state configuration exhibiting a triaxial deformation of $\gamma \sim30^\circ$. These results are compared with a recent version of the triaxial rotor model. It is worth noting that the nucleus $^{72}$Ge and its neighbors have also been the focus of detailed and varied spectroscopic investigations using multi-nucleon transfer reactions: these are summarized with an attempt at a simple pairing occupancy picture in Ref.~\cite{RevModPhys.83.1467}, but this picture has not been connected to the quadrupole collective properties of the nuclei involved.

\begin{center}
\begin{figure}[]
\hspace{-1.em}
\includegraphics[width=0.85\columnwidth]{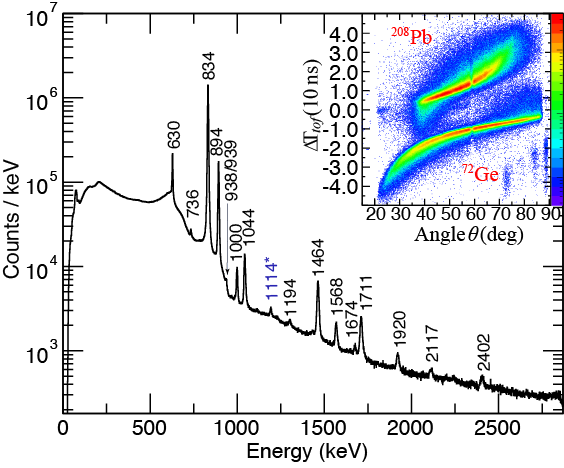} %0.49
\caption{\label{fig:specall} (Color online) Doppler-corrected $\gamma$-ray spectrum obtained in kinematic coincidence with $^{72}$Ge ions. Note that the 1114-keV transition belongs to $^{72}$Ge, but has not been placed and, thus, was not included in the analysis. The insert illustrates the performance of the CHICO2 array in discriminating between the projectile and target nuclei.} \end{figure}
\end{center}

\begin{center}
\begin{figure}[t!]
\includegraphics[width=\columnwidth]{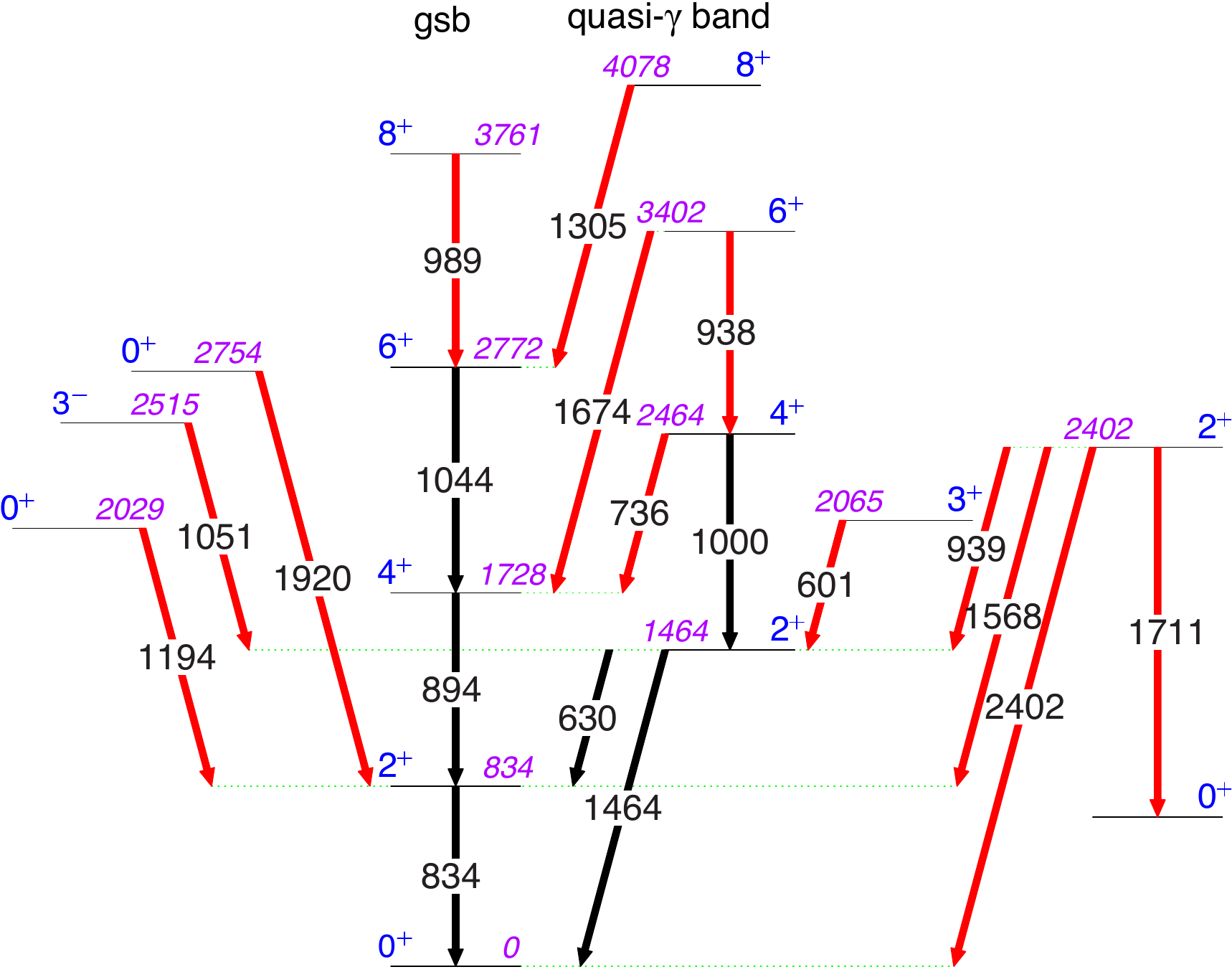}
\caption{\label{fig:levelsch} (Color online) A partial level scheme showing all the relevant levels (black + red) identified in the present Coulomb excitation measurement. Transitions in black (6 in total) are those observed in the most recent measurement~\cite{Kot1990646}; all others have been seen here for the first time in Coulomb excitation, but had been seen and established in other types of measurements~\cite{Abriola20101}}.
\end{figure}
\end{center}

Multi-step Coulomb excitation of the $^{72}$Ge nucleus was carried out by bombarding a $0.5$ mg/cm$^2$-thick $^{208}$Pb target, sandwiched between a 6 $\mu \mathrm{g}/\mathrm{cm}^2$  Al front layer and a 40  $\mu \mathrm{g}/\mathrm{cm}^2$ C backing, with a 301-MeV $^{72}$Ge beam delivered by the ATLAS facility at Argonne National Laboratory. 
The $\gamma$ rays emitted in the deexcitation were detected with the $\gamma$-ray tracking array, GRETINA~\cite{Paschalis201344}. At the time of the experiment, this array consisted of 28 highly-segmented coaxial high-purity germanium (HPGe) crystals grouped into 7 modules. The scattered projectile and recoiling target nuclei  were detected in kinematic coincidence with the $\gamma$ rays by CHICO2, an array of 20 position-sensitive parallel-plate avalanche counters arranged symmetrically around the beam axis~\cite{chico2}. CHICO2 covers $68\%$ of the solid angle around the target and provides a position resolution (FWHM) better than $1.6^{\circ}$ in $\theta$ (polar angle) and $2.5^{\circ}$ in $\phi$ (azimuthal angle) relative to the beam axis. In addition, it achieves a time resolution of $1.2$ ns (FWHM), sufficient for a measurement of the time-of-flight difference, $\Delta T_{tof}$, between the reaction products as a function of the scattering angle, $\theta$. This enables an event-by-event reconstruction of the reaction kinematics and a precise Doppler-shift correction of the $\gamma$ rays. Figure~\ref{fig:specall} provides the resultant $\gamma$-ray spectrum measured in coincidence with the scattered $^{72}$Ge projectiles. The insert depicts a two-dimensional histogram of $\Delta T_{\mathrm{tof}}$ vs. $\theta$ demonstrating the clear separation (with a mass resolution of $\Delta m/m \approx$ $5\%$) between the reaction participants. A partial scheme, incorporating all the relevant levels populated in the present measurement, is provided in Fig.~\ref{fig:levelsch}, where the black-colored transitions are those observed in the prior Coulomb-excitation measurement~\cite{Kot1990646}. The resolving power and efficiency of the present experiment is illustrated by the many additional transitions given in red.  

\vspace{-1.5cm}

The Coulomb-excitation analysis was performed using the coupled-channel  least-squares search code, \textsc{gosia}~\cite{gosia, gosia2}, which constructs a standard $\chi^2$ function based on comparisons between the measured $\gamma$-ray yields and theoretical ones calculated from an initial set of transition and static matrix elements. Since the Coulomb cross sections depend on the relative phases as well as on the sign and magnitude of the $E\lambda$ matrix elements (ME), a random set of the MEs used as starting values in the $\chi^2$ search were chosen to sample all possible signs of the interference term. Furthermore, all intra-band transitions within the ground-state and quasi-$\gamma$ bands as well as the $2^+_3 \rightarrow 0^+_2$ transition were assigned positive MEs. The signs of all other MEs for the inter-band transitions were determined relative to these, and varied during the fitting process to avoid being trapped in local $\chi^2$ minima. It should be noted that the signs of these MEs, i.e., intra- and inter-band transition matrix elements are not experimental observables. In contrast, the signs of the static or diagonal matrix elements determine whether the deformation is prolate (negative ME for $K=0$) or oblate (positive ME for $K=0$). 

In order to enhance the sensitivity to the MEs and to exploit the dependence of the excitation probability on the particle scattering angle, the data were partitioned into seven angular subsets: $31^\circ - 40^\circ$, $41^\circ - 50^\circ$, $51^\circ - 60^\circ$, $61^\circ - 70^\circ$, $71^\circ - 85^\circ$, $96^\circ - 130^\circ$, and $131^\circ - 165^\circ$. This resulted in a total of about 70 data points for the $\chi^2$ analysis. In addition, known spectroscopic data  such as lifetimes, branching and $E2/M1$ mixing ratios were included as constraints of the relevant parameters during the fitting process. The final MEs and their associated errors are displayed in Table~\ref{tab2}. For the purpose of this discussion, only the relevant $E2$ MEs are tabulated, while the $E3$ ME for the $3^-_1$ state is given here; {\it i.e.}, $\bra{I_i}|M(E3)|\ket{I_f} = 0.199(4)\; e\mathrm{b}^{3/2}$. The quoted errors for all MEs were derived in the standard way by constructing a probability distribution in the space of fitted parameters and requesting the total probability to be equal to the chosen confidence limit (in this case 68.3\%). These errors include the statistical and systematic contributions as well as those arising from cross-correlation effects.

\begingroup
\squeezetable
\begin{table}
\caption{\label{tab2} Reduced $E2$ matrix elements for transitions of $^{72}$Ge, deduced from the present work, in comparison with previous measurements.}
\ra{1.1}
\begin{tabular}{rrrrrrrrr}
\hline\hline
\multicolumn{1}{c}{\multirow{2}{*}{$I_i^\pi \rightarrow I_f^\pi$}} & \multicolumn{6}{c}{$\bra{I_i} |\mathcal{M}(E2)| \ket{I_f}$ $(e\mathrm{b})$}                                                                                                                                       \\ \cline{3-7} 
\multicolumn{1}{c}{}                     && \multicolumn{1}{c}{This Work}     && \multicolumn{1}{c}{Ref.~\cite{Kot1990646}} && \multicolumn{1}{c}{Refs.~\cite{PhysRevC.22.2420,PhysRevC.22.1530}} \\ \hline
$0_1^+\rightarrow 2_1^+$		&&	0.457(4)	&&	0.46(1)		&&	0.457(7) \\
$2_1^+\rightarrow 4_1^+$		&&	0.90(2)	&&	0.89(4)		&&	0.76(4)   \\
$4_1^+\rightarrow 6_1^+$		&&	$1.11_{-0.05}^{+0.04}$	&&	1.2(3)		&&	 	       \\
$6_1^+\rightarrow 8_1^+$		&&	$1.1_{-1.6}^{+0.2}$	&&				&&	 		\\            
$2_1^+\rightarrow 2_1^+$		&&   $-0.16_{-0.02}^{+0.07}$	&&	$-0.16_{-0.07}^{+0.10}$ &&$-0.17(8)$\\
$4_1^+\rightarrow 4_1^+$		&&   $-0.14_{-0.04}^{+0.09}$	&&	$-0.01(1)$		&&	 		\\
$6_1^+\rightarrow 6_1^+$		&&   $-0.20_{-0.25}^{+0.08}$	&&	$-0.1(5)$		&&	        		\\   
$2_1^+\rightarrow 0_2^+$		&&  $0.35_{-0.02}^{+0.01}$	&&	$0.36(4)$		&&	$|0.45(2)|$       \\ 
$4_1^+\rightarrow 2_2^+$		&&  $-0.06_{-0.04}^{+0.03}$ 			&&	$-0.08(5)$		&&	\\
$6_1^+\rightarrow 4_2^+$		&& $0.28_{-0.05}^{+0.10}$	&&	$<\bigl|0.4\bigr|$		&&				\\

\hline
$2_2^+\rightarrow 3_1^+$		&& $1.19(2)$	&&		&&				\\
$2_2^+\rightarrow 4_2^+$		&&   	$0.58_{-0.01}^{+0.05}$	&&	$0.41^{+0.06}_{-0.02}$		&&	 		\\
$4_2^+\rightarrow 6_2^+$		&&  0.74(2)	&&		&&						\\  
$2_2^+\rightarrow 2_2^+$		&&   	$0.179_{-0.06}^{+0.03}$	&&	$0.30(10)$	&&	 			\\
$3_1^+\rightarrow 3_1^+$		&& $\bigl|0.001\pm0.521\bigr|$  	&&			&&				\\ 
$4_2^+\rightarrow 4_2^+$		&&   	$-0.29^{+0.04}_{-0.20}$		&&	$<\bigl|0.4\bigr|$		&&	 		\\
$0_1^+\rightarrow 2_2^+$		&& 0.030(1)	&&	$0.034(5)$	&&	$\bigl|0.031(7)\bigr|$	\\
$0_2^+\rightarrow 2_2^+$		&&   	$0.0144(6)$		&&	$\bigl|0.019(5)\bigr|$	&&	 0.016(3)\\
$2_1^+\rightarrow 2_2^+$		&& $0.65_{-0.02}^{+0.01}$	&&	$0.78(2)$		&&	$|0.75(8)|$	\\ 
$2_1^+\rightarrow 4_2^+$		&&   	$0.035(6)$	&&	$\bigl|0.024(5)\bigr|$      		&&	 		\\
$4_1^+\rightarrow 4_2^+$		&&  $0.43(10)$	&&	$0.60(10)$	&&				\\ 
$4_1^+\rightarrow 6_2^+$		&&  $0.178_{-0.007}^{+0.020}$	&&		&&			                 	\\ 
$6_1^+\rightarrow 6_2^+$		&&  $0.18_{-0.08}^{+0.25}$	&&		&&						\\
$6_1^+\rightarrow 8_2^+$		&&  $0.23_{-0.10}^{+0.05}$					&&		&&						\\
\hline
$0_1^+\rightarrow 2_3^+$		&& $0.044(1)$  			&&	$\le$$\bigl|0.022 \bigr|$		&&\\
$0_2^+\rightarrow 2_3^+$		&& $0.279_{-0.004}^{+0.002}$	&&	$\le$$\bigl|0.13 \bigr|$		&&			      \\
$2_1^+\rightarrow 2_3^+$		&& $0.243_{-0.004}^{+0.002}$	&&	$\le$$\bigl|0.12 \bigr|$		&&			     \\
$2_2^+\rightarrow 2_3^+$		&& $0.49_{-0.01}^{+0.02}$  	&&	$\le$$\bigl|0.32 \bigr|$		&&	         	\\
$2_3^+\rightarrow 2_3^+$		&& $-0.02_{-0.14}^{+0.03}$  	&&			&&			\\
\hline \hline
\end{tabular}
\end{table}
\endgroup

\paragraph*{}

\begin{figure}[t!]
\begin{center}
\includegraphics[width=0.85\columnwidth]{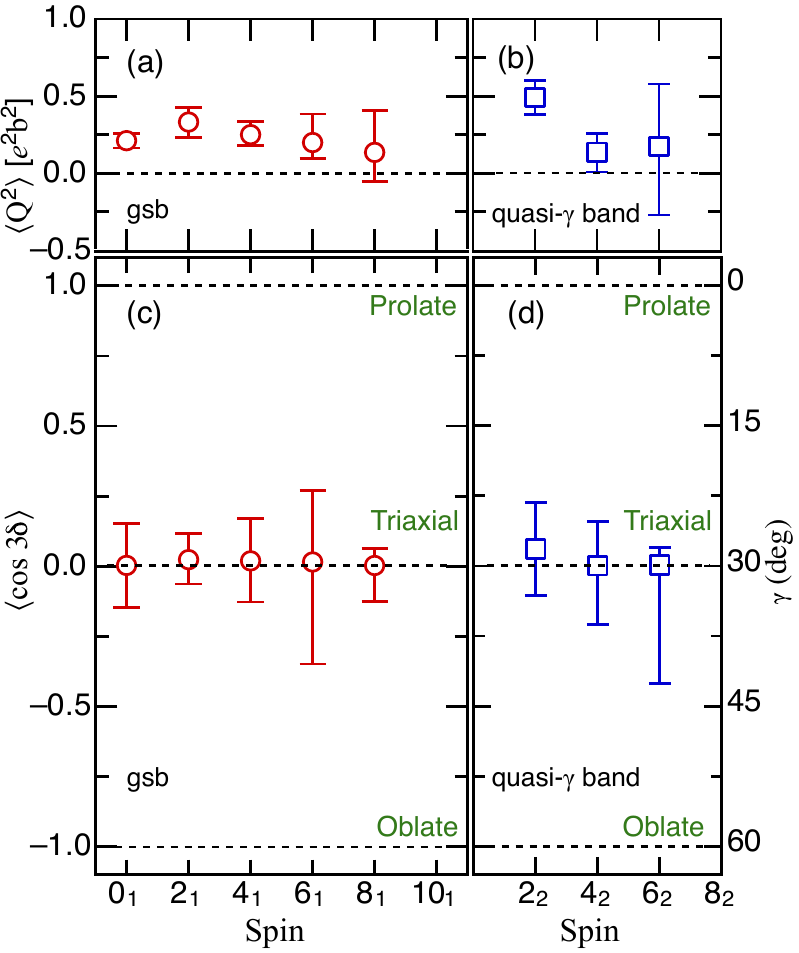}
\caption{(Color online) $\left < Q^2 \right >$ (top) and $\left < \mathrm{cos}\; 3\delta \right > $ quantities (bottom) obtained from the invariant sum rule analysis described in the text. Limits for the various shapes are indicated by the dashed lines.}
\label{q2cos3d}
\end{center}
\end{figure}

The quadrupole collectivity of the low-lying states in $^{72}$Ge and the associated shapes can be analyzed quantitatively using the model-independent invariant sum rules of Kumar~\cite{KK1} and Cline~\cite{KK2}, which construct the deformation parameters [$\left < Q^2 \right >$, $\left < \mathrm{cos}\; 3\delta \right >$] from a complete set of $E2$ MEs determined via Coulomb excitation. The quadrupole invariants, $\left < Q^2 \right >$ and $\left < \mathrm{cos}\; 3\delta \right >$, describe the nuclear charge distribution via rotationally-invariant scalar products of the quadrupole operators which relate the reduced $E2$ MEs to the quadrupole deformation parameters~\cite{KK2}. In this formalism, the parameters, $Q$ and $\delta$, have correspondence to the collective model variables, $\beta$ and $\gamma$, defining the overall quadrupole deformation and axial asymmetry, respectively. The experimentally determined expectation values of the sum of products of $E2$ MEs, $\left < Q^2 \right >$, and the axial asymmetry quantity, $\left <\mathrm{cos}\; 3\delta \right >$, for the relevant states within the ground-state (gsb) and the quasi-$\gamma$ bands are presented in Fig.~\ref{q2cos3d}. The almost constant and nonzero value of $\left < Q^2 \right >$ [Fig.~\ref{q2cos3d}(a)] indicates that $^{72}$Ge is deformed in its ground-state band. In addition, the behavior of the asymmetry parameter [Fig.~\ref{q2cos3d} (c)] is consistent with a triaxially-deformed shape for this band. Figs.~\ref{q2cos3d}(b) and \ref{q2cos3d}(d) provide evidence that a similar conclusion can be drawn for the quasi-$\gamma$ band as well. 

\begin{figure}[!t]
\begin{center}
\includegraphics[width=0.85\columnwidth]{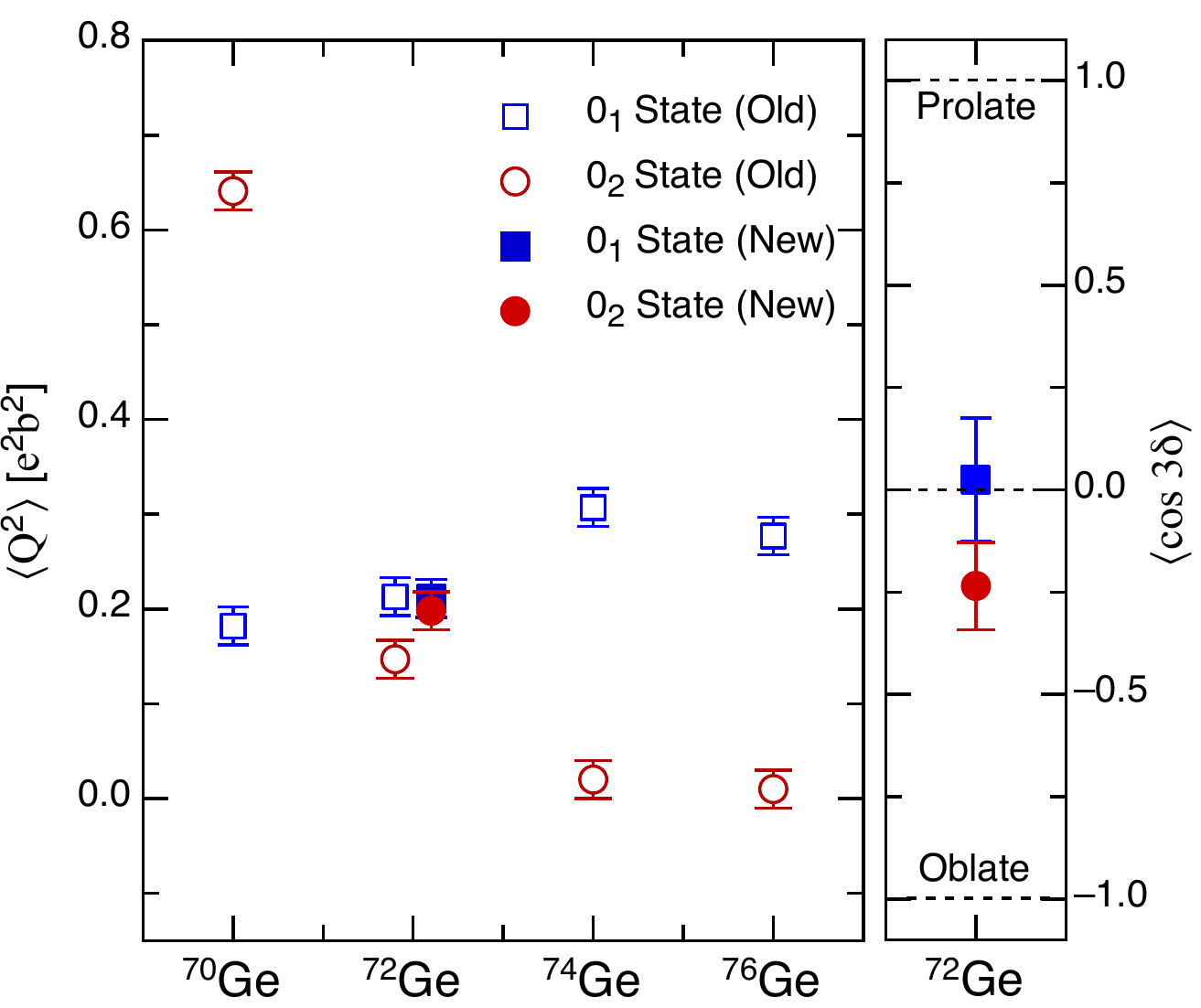}
\caption{(Color online) $\left<Q^2\right> $ and $\left < \mathrm{cos}\; 3\delta \right >$ quantities for the $0_1^+$ and $0_2^+$ states in the Ge isotopes. The `old' data (open symbols) are taken from Ref.~\cite{RevModPhys.83.1467}.}
\label{q2_0p}
\end{center}
\end{figure}

Figure~\ref{q2_0p} presents the invariant sum rules analysis for the two lowest $0^+$ states of $^{72}$Ge, in comparison with those of neighboring Ge isotopes. The present and the previous results are given with filled and open symbols, respectively. While it is evident from Fig.~\ref{q2_0p} that a well-defined shape transition occurs when going from mass 70 to 72; {\it i.e.}, the ground-state configuration for $^{70}$Ge becomes the $0^+_2$ excitation in $^{74,76}$Ge and vice versa for the $0^+_1$ levels in the latter two nuclei, the results of the present investigation (filled symbols) indicate a more subtle transition and show that the $0_1^+$ and $0_2^+$ states in $^{72}$Ge  have essentially the same $\left<Q^2\right>$ value. This is in contrast to previous investigations (see open symbols for $^{72}$Ge), partly because of the $0_2^+ \rightarrow 2_3^+$ ME, which was determined here for the first time, and found to contribute significantly to the sum rule. It should be noted that, while the mean value of the shape parameter is almost the same for these two $0^+$ states, it does not necessarily follow that they are characterized by the same shape. In fact, the identical magnitude of the  $\left<Q^2\right>$ values can be interpreted as a consequence of mixing of two distinct (or unperturbed) configurations characterizing these states, each being associated with a different shape. Within a simple two-state mixing model, the $0_1^+$ and $0_2^+$ states can be expressed as linear combinations of two unmixed $0^+_{u1,u2}$ basis levels with different mixing amplitudes and unique deformations~\cite{PhysRevC.75.054313}. Assuming no interaction between the intrinsic states; {\it i.e.}, $\bra{I_{u1}}|M(E2)|\ket{I_{u2}}=0$, and using the four MEs connecting the $0^+_{1,2}$ to the $2^+_{1,3}$ mixed states, the intrinsic, unmixed basis states, initially separated by $\Delta_{int} = 32$ keV, are shifted by 330 keV with respect to one another by a mixing ME of 345 keV. The wavefunction of the mixed $0_1^+$ state contains an amplitude $\mathrm{cos}\theta_0 = 0.72(3)$ of the unmixed ground-state wavefunction, which indicates maximum mixing [$\mathrm{cos}^2\theta_0 = 0.52(4)$], and is consistent with values derived in two-neutron and alpha transfer measurements~\cite{vandenBerg1982239}. In principle, the mixing strength can also be determined on the basis of the known E0 monopole strength  between the two $0^+$ states [$\rho^2(E0; 0^+_2 \rightarrow 0^+_1) \times 10^3= 9.18(2)$ in $^{72}$Ge]~\cite{Kibedi200577}. This would however, require knowledge of the difference in the mean-square charge radii of the two configurations which is at present unavailable. Thus, while the simple two-state mixing calculations describe the mixing between configurations of the $0^+$ states reasonably well, ambiguities associated with the influence of the highly-mixed $2^+$ ``subspace" (comprising the $2^+_1$, $2^+_2$, and $2^+_3$ states), as evidenced by the various decay branches and magnitudes of the linking MEs (see Tab.~\ref{tab2}), remain to be addressed. Hence, a complete description requires, at the minimum, a three-band mixing calculation that takes into account the contributions of the $2^+$ states with, in addition, the inclusion of triaxiality, as indicated by the $\left < \mathrm{cos}\; 3\delta \right >$ analysis of Fig.~\ref{q2_0p}.

\begin{figure*}[t!]
\begin{center}
\includegraphics[width=1.5\columnwidth]{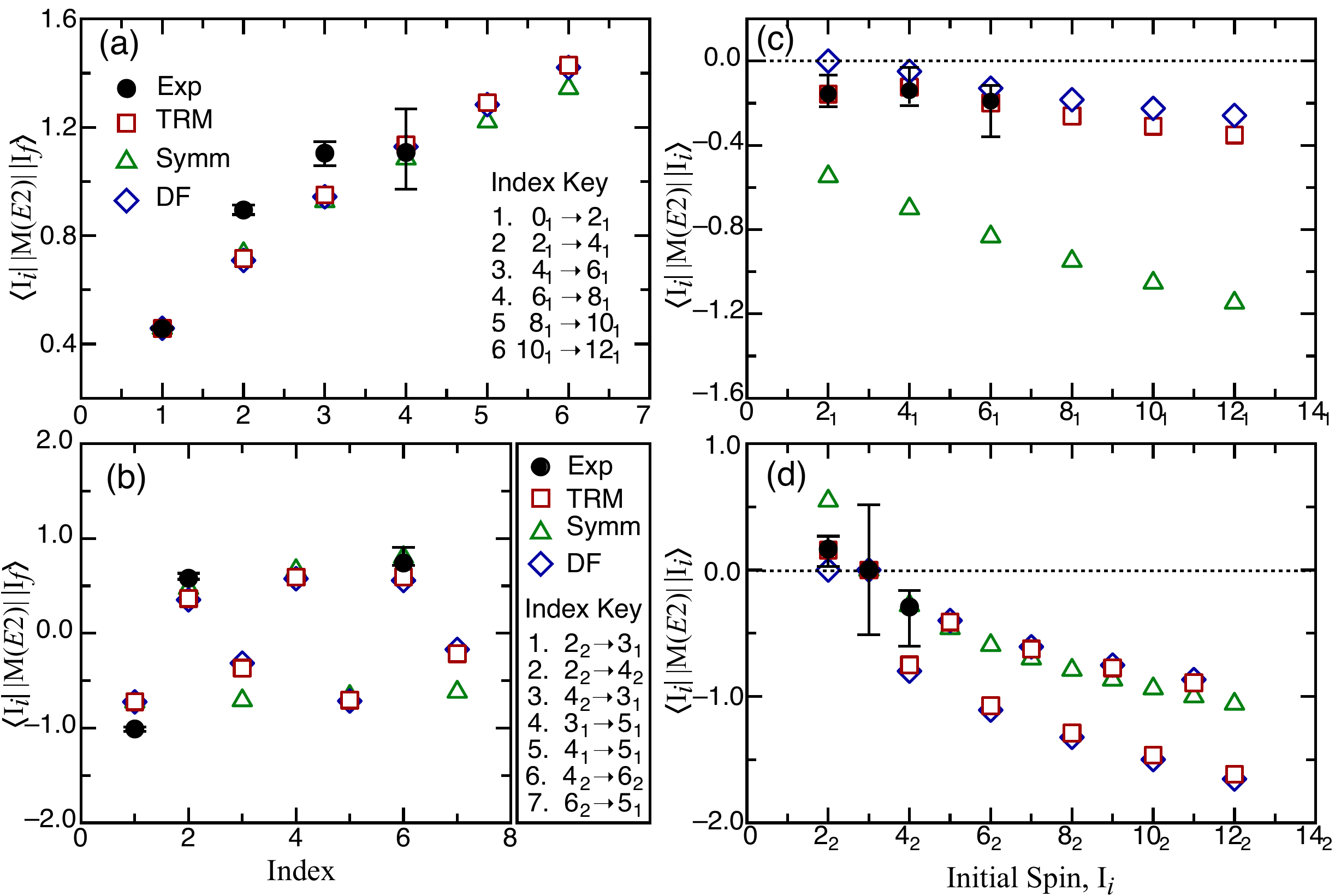}
\caption{(Color online) (a) and (b): Experimental transition matrix elements for intra-band transitions, $\bra{I_i}|M(E2)|\ket{I_f}$ in comparison with theoretical calculations using the triaxial rotor model (TRM), symmetric (symm), and Davydov-Filipov (DF) models. (c) and (d): Similar comparisons for static quadrupole moments  $\bra{I_i}|M(E2)|\ket{I_i}$ in the ground and quasi-$\gamma$ bands.}
\label{gb-gm-me}
\end{center}
\end{figure*}

In order to quantify the role of triaxiality and present a better understanding of shape coexistence in the low-lying states, a revised version of the triaxial rotor model with independent inertia and electric quadrupole tensors \cite{PhysRevC.70.024308, WDK06, JMA08, JMA09, JMA10} was applied to the newly deduced $E2$ MEs of $^{72}$Ge. This version is a departure from the standard use of irrotational flow moments of inertia; {\it e.g.}, the Davydov-Filippov rotor model \cite{DAV}. Using this model, the $E2$ MEs for states within the ground and quasi-$\gamma$ bands were calculated \cite{JMA06a} with a minimum set of assumptions and compared to the experimental results. The three model parameters required to describe the $E2$ MEs of a triaxial rotor include the deformation, $Q_0$, the asymmetry of the electric quadrupole tensor, $\gamma$, and the mixing angle of the inertia tensor, $\Gamma$. These parameters are determined analytically in this study from the $\reduceE{0_1}{2_1}$, $\reduceE{0_1}{2_2}$, and $\reduceE{2_1}{2_1}$ experimental MEs (cf.~Ref.~\cite{JMA08}) and result in $Q_0=1.45$~eb, $\gamma=27.0^\circ$, and $\Gamma=-23.2^\circ$.

\begin{figure}[b]
\begin{center}
\includegraphics[width=0.75\columnwidth]{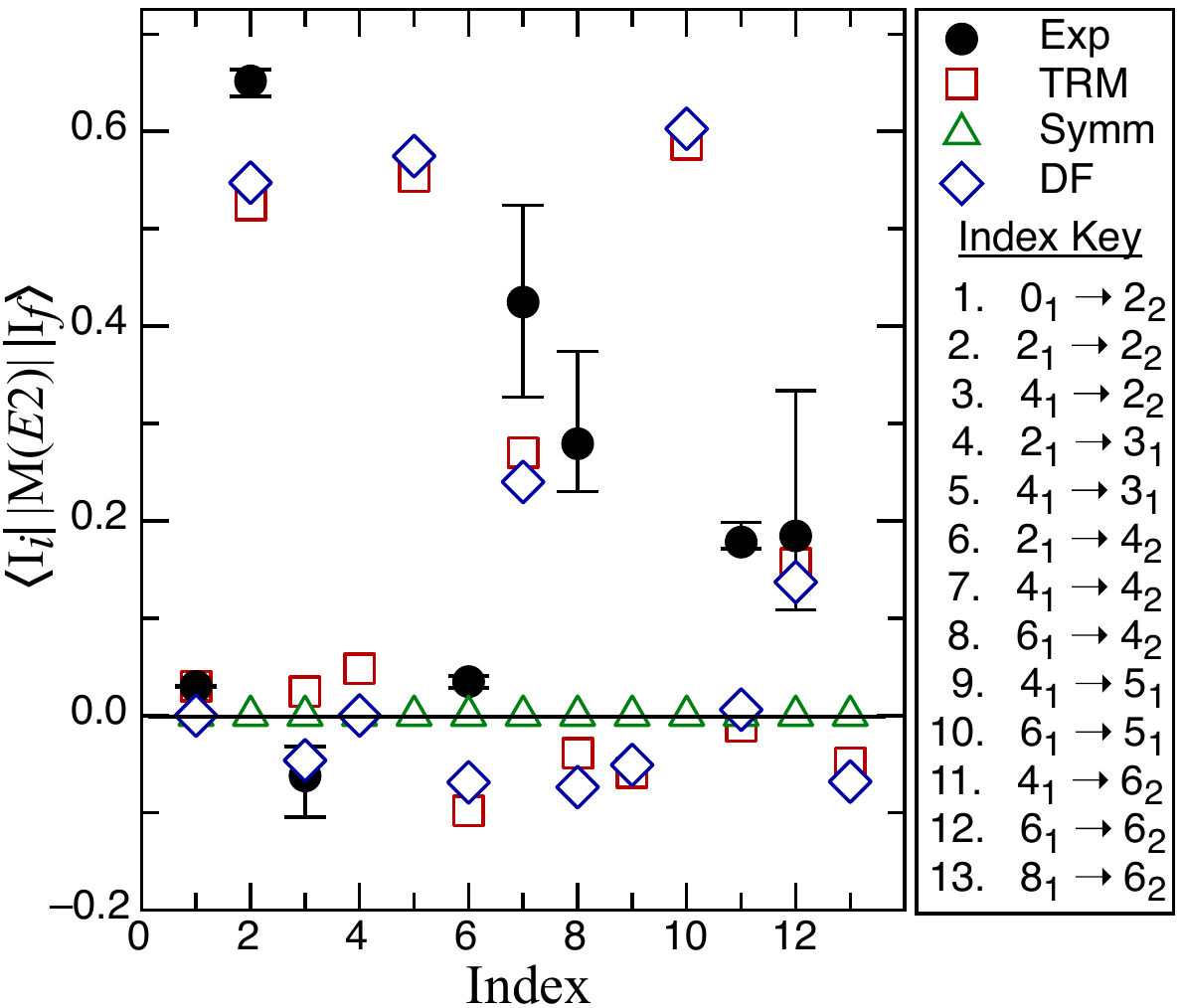}
\caption{(Color online) Experimental transition matrix elements for inter-band transitions in comparison with theoretical calculations using the triaxial rotor model (TRM), symmetric rotor (symm) and the Davydov-Filipov (DF) models.}
\label{gb-gm-links}
\end{center}
\end{figure}

The results of triaxial rotor model calculations \cite{JMA06a}, designated as (TRM), are compared with the data in Figs.~\ref{gb-gm-me} and \ref{gb-gm-links}. For completeness, calculations for a symmetric rotor (Symm) and triaxial rotor with irrotational flow moments of inertia (DF) \cite{DAV} are also included. All three versions of the rotor model equally reproduce the ground [Fig.~\ref{gb-gm-me}(a)] and quasi-$\gamma$ intra-band transitions [Fig.~\ref{gb-gm-me}(b)]. Beyond this point, however, the symmetric rotor (Symm) has little to no value in describing the data. For the quasi-$\gamma$ to ground inter-band transitions, the TRM and DF calculations appear to perform equally well, predicting large values when the experimental values are large and likewise when they are small (see Fig.~\ref{gb-gm-links}). However, there is a consistent failure to reproduce the $\reduceE{6_1}{4_2}$ and $\reduceE{4_1}{6_2}$ MEs (see indices 8 and 11 in Fig~\ref{gb-gm-links}); these $\Delta I=2, \Delta K=2$ transitions have been shown to be very sensitive to interference effects \cite{JMA08, JMA10}. With the exception of the $4_2$ state, perhaps the most important outcome is the ability of the TRM calculations to reproduce the static quadrupole moments; {\it i.e.}, the diagonal $E2$ MEs $\reduceE{I_i}{I_i}$. This is presented in Figs.~\ref{gb-gm-me}(c) and \ref{gb-gm-me}(d), where both the experimental $\reduceE{I_i}{I_i}$ values and the general trend with spin are in good agreement with expectations of a triaxial rotor.

Configuration mixing between two triaxial rotor models was utilized \cite{JMA06b} to investigate further the nature of the first excited $0_2^+$ level and the $2_3^+$ state possibly built upon it, and to determine their impact on the $E2$ MEs of the ground and quasi-$\gamma$ bands. A simple approach was adopted to extract the model parameters without any effort to further adjust the parameters for better agreement. This approach was to factorize the solution into two $2\times2$ subspaces in order to solve the $\Delta K=2$ triaxial mixing (see above) and $\Delta K=0$ configuration mixing. Then, using the parameters from these solutions, the $E2$ MEs for the full $4\times4$ space were calculated. The $\Delta K=0$ configuration mixing was solved using the $\reduceE{0_1}{2_1}$, $\reduceE{2_1}{0_2}$, $\reduceE{0_1}{2_3}$, and $\reduceE{0_2}{2_3}$ experimental MEs.  The adopted parameters for the configuration mixing are $Q_{0_1}=1.93$~b, $Q_{0_2}=0.57$~b, $\theta_{I=0,\Delta K=0}=43.2^\circ$, and $\theta_{I=2,\Delta K=0}=21.1^\circ$. From the $\Delta K=0$ configuration mixing alone, the MEs $\reduceE{2_1}{2_1}$, $\reduceE{2_3}{2_3}$, and $\reduceE{2_1}{2_3}$ are predicted to be 0.66, -0.28, and -0.17~$e$b, respectively, but the corresponding experimental values are -0.16, -0.02, and 0.24~$e$b. Hence, $\Delta K=0$ configuration mixing alone is not sufficient to describe the data and suggest including $\Delta K =2$ triaxial mixing.

The triaxial and configuration mixing solutions were combined to give the following overall adopted parameters: $Q_{0_1}=1.93$~b, $\gamma_1=27^\circ$, and $\Gamma_1=-23.2^\circ$ for the first triaxial rotor, $Q_{0_2}=0.57$~b, $\gamma_2=30^\circ$, and $\Gamma_2=-30^\circ$ for the second one ($30^\circ$ was assumed for simplicity), and $\theta_{I=0,\Delta K=0}=43.2^\circ$ and $\theta_{I=2,\Delta K=0}=21.1^\circ$ for the configuration mixing. This approximation scheme results in calculated $E2$ MEs that are typically within a few percent of the exact solution \cite{JMA06b}. 

\begin{figure}[b!]
\begin{center}
\includegraphics[width=0.8\columnwidth]{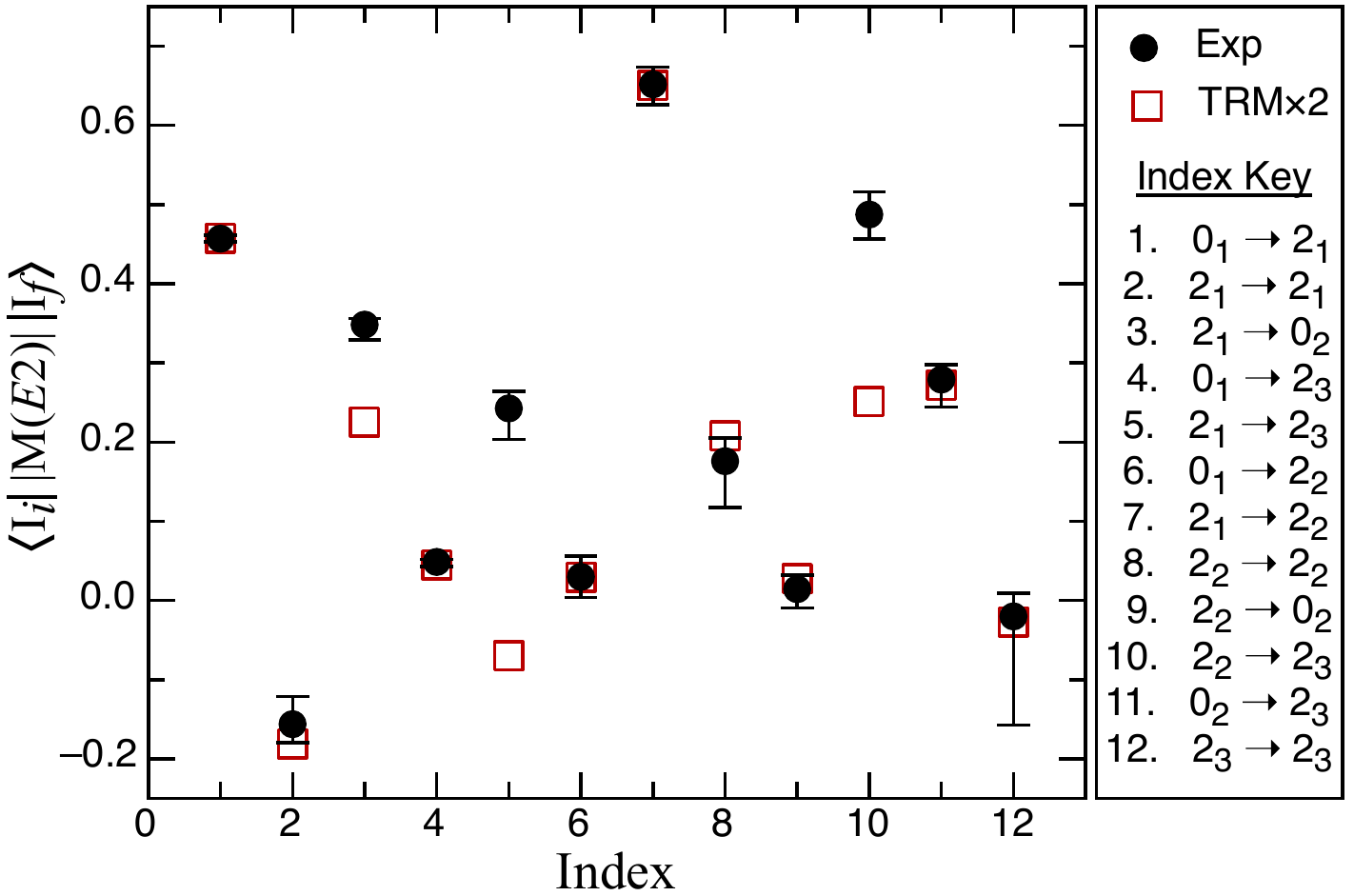}
\caption{(Color online) Experimental matrix elements in comparison with triaxial-triaxial mixing calculations using the triaxial rotor model with independent moments of inertia.}
\label{tri-tri-mixing}
\end{center}
\end{figure}

The results of configuration mixing calculations with the two-triaxial-rotor model~\cite{JMA06b}, designated as (TRM$\times$2), are compared with the experimental values in Fig.~\ref{tri-tri-mixing}. The diagonal $E2$ MEs $\reduceE{2_1}{2_1}$, $\reduceE{2_2}{2_2}$, and $\reduceE{2_3}{2_3}$ ({\it i.e.}, quadrupole moments, corresponding to indices 2, 8, and 12, respectively, in Fig.~\ref{tri-tri-mixing}) are well described. However, the $\reduceE{2_1}{2_3}$ ME (index 5) is still not reproduced; nor is $\reduceE{2_2}{2_3}$ (index 10). Interestingly, the $\reduceE{2_1}{2_2}$ and $\reduceE{2_2}{0_2}$ MEs are predicted well, and configuration mixing substantially improves the description of the $\reduceE{2_1}{0_2}$ ME (index 3) over that from a single triaxial rotor (see above). 
It can be concluded that this configuration mixing is well justified in describing the $E2$ MEs between the $0_1$, $0_2$, $2_1$, and $2_2$ states, which can also be described with only four parameters; {\it i.e.},  $Q_{0a}=1.93$~b, $\gamma_a=27^\circ$, and $\Gamma_a=-23.2^\circ$ for the first triaxial rotor and $\theta_{I=0,\Delta K=0}=43.2^\circ$ for configuration mixing with an isolated $0^+$ state. However, the evidence is not so compelling for the $2_3^+$ level. This state could possibly be viewed as a member of a two-phonon $\gamma$ vibration, but this would necessitate the presence of  excited $K=0,4$ bands at nearly twice the energy of the first excited $K=2$ sequence. There is at present no evidence for this conjecture. 

Finally, the sum of $2^+$ quadrupole moments provides another interesting perspective on triaxiality and shape coexistence. This sum over a complete basis or subspace should be zero \cite{JMA13}. For a triaxial-rotor model, there are only two $2^+$ states, whose quadrupole moments must be equal and opposite. For \element{72}{Ge}, $\reduceE{2_1}{2_1}=-0.16(^{+7}_{-2})$ and $\reduceE{2_2}{2_2}=0.179(^{+3}_{-6})$, yielding a sum of $0.02(^{+8}_{-6})$. If the sum is made over the three $2^+$ states observed in the present study, where $\reduceE{2_3}{2_3}=-0.02(^{+2}_{-14})$, the result is $0.00(^{+8}_{-15})$. Assuming  that the $2_3^+$ state is part of a second triaxial rotor configuration, there must then be another $2^+$ level with a diagonal $E2$ ME equal to $0.00(^{+8}_{-15})$, consistent with $\gamma_2=30^\circ$. It can thus be concluded that the quadrupole moment sum is consistent with shape coexistence between two triaxial structures, but it is also consistent with other scenarios; {\it e.g.}, shape coexistence between triaxial and spherical structures. It can also be argued, based on the magnitude and the general pattern of the MEs linking the $2^+_3$ level to the lower-lying states, that this state may not be directly associated with the $0^+_2$ level, as has often been assumed. In fact, the TRM calculations account for only $\sim32\%$ of the measured $B(E2)$ value for this state. The introduction of gamma softness, perhaps resulting from a two-phonon character, may be required to account for the remaining strength. Further work will be required to clarify this issue.

In summary, the collective properties of the ground-state and quasi-$\gamma$ bands as well as the first excited $0^+$ and $2^+_3$ states in $^{72}$Ge have been investigated in a model-independent way using sub-barrier multi-step Coulomb excitation. The model-independent shape invariants obtained from the Kumar-Cline sum rule analysis of the low-spin structure provide compelling evidence for the coexistence of two triaxially-deformed configurations associated with the $0^+_1$ and $0^+_2$ states. 
Within a two-state mixing model, the extracted matrix elements agree with this shape coexistence interpretation, but also require maximum mixing between the wavefunctions of the first two $0^+$ states. 
The results of multi-state mixing calculations performed within the framework of the triaxial rotor model demonstrate the importance of the triaxial degree of freedom and indicate that the low-spin structure of $^{72}$Ge can be adequately described by mixing of two triaxial rotors. This result reveals that a nucleus, apparently unlike any other known collective one, possesses a simple underlying structure. This interpretation, by way of coexistence of different shapes, supports structural simplicity when the phenomenology appears complex.

\paragraph*{}
This material is based upon work supported by the U.S. Department of Energy, Office of Science, Office of Nuclear Physics under Contract numbers DE-AC02-06CH11357 and DE-AC52-07NA27344, and under Grant Numbers DE-FG02-94ER40834 and DE-FG02-08ER41556, and by  the National Science Foundation. This research used resources of ANL's ATLAS facility, which is a DOE Office of Science User Facility.

\bibliographystyle{apsrev4-1}
\bibliography{72Ge-paper}
\end{document}